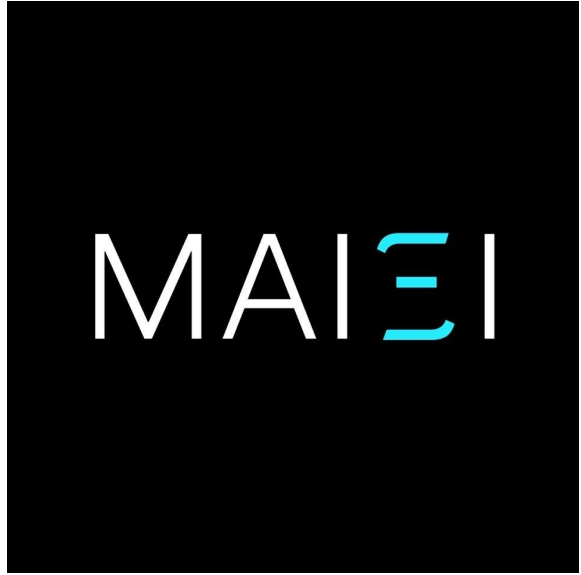

Montreal AI Ethics Institute

*An international, non-profit research institute helping humanity define its place in a world increasingly driven and characterized by algorithms*

Website: https://montrealethics.ai
Newsletter: https://aiethics.substack.com

# SECure: A Social and Environmental Certificate for AI Systems

Abhishek Gupta[1,2], Camylle Lanteigne[1,3], and Sara Kingsley[4]

[1] Montreal AI Ethics Institute
[2] Microsoft
[3] McGill University
[4] Carnegie Mellon University

**Abstract**

*In a world increasingly dominated by AI applications, an understudied aspect is the carbon and social footprint of these power-hungry algorithms that require copious computation and a trove of data for training and prediction. While profitable in the short-term, these practices are unsustainable and socially extractive from both a data-use and energy-use perspective. This work proposes an ESG-inspired framework combining socio-technical measures to build eco-socially responsible AI systems. The framework has four pillars: compute-efficient machine learning, federated learning, data sovereignty, and a LEEDesque certificate.*

*Compute-efficient machine learning is the use of compressed network architectures that show marginal decreases in accuracy. Federated learning augments the first pillar's impact through the use of techniques that distribute computational loads across idle capacity on devices. This is paired with the third pillar of data sovereignty to ensure the privacy of user data via techniques like use-based privacy and differential privacy. The final pillar ties all these factors together and certifies products and services in a standardized manner on their environmental and social impacts, allowing consumers to align their purchase with their values.*

**Introduction and background**

This research project aims to take the most comprehensive approach to assess the environmental and social impacts of machine learning. Current machine learning practices emit excessively large amounts of carbon dioxide, while also requiring access to expensive and highly specialized hardware (Strubell et al, 2019). Additionally, issues related to data privacy abound. We employ an 'environment, society, and governance' (ESG) framework to understand—and subsequently shift—how machine learning and the actors behind its development affect our world. Through its four pillars, our ESG framework targets researchers, industry, and consumers.

Our motivation for embarking on this work is to surface the tradeoffs that researchers and practitioners should consider when they develop more complex models which call for the use of larger datasets. These efforts can yield more predictive power, but do not come without a second-order effect, which is often abstracted away from developers. This is even more so the case for consumers of these systems who do not see the environmental and social impacts as AI is integrated into existing software systems as an additional capability. Through this work, we seek to bring these impacts to the foreground and provide the tools and necessary data for both consumers and developers to make informed decisions when considering the use of AI systems. There are already discussions, especially as it relates to the use of supervised machine learning techniques that require labelled training datasets and how that has led to the emergence of a shadow workforce that toils in the background (Brawley & Pury, 2016) enabling some of the impressive feats we have seen these systems accomplish in recent times.

This brings up issues of unjust labor practices and unfair compensation that lie behind some of the modern conveniences that are powered by AI-enabled solutions. It calls for a greater scrutiny on the entire supply chain of building and deploying these systems such that people are able to make choices that are akin to "fair-trade" product labels guiding consumers on some of the practices that were involved in the generation of the products and services. Ultimately, having empowered users who better understand the impacts of their purchasing decisions (Hainmueller et al., 2015) can become a powerful lever for galvanizing change in the current practices in the development and deployment of machine learning systems.

**Existing and related work**

The work done at the intersection of AI and environmental impact is still very sparse. Most work that engages with both of these topics addresses ways in which AI can help counteract or adapt to the impacts of climate change. This work is unquestionably extremely valuable, and we commend researchers for attempting to use AI in such a way. We believe it is, as a result, all the more important to engage with how AI is unnecessarily and excessively environmentally harmful.

Crucially, many of the elements that make the field of machine learning carbon intensive can also make it inaccessible. For instance, the need for onerous hardware and extremely large compute resources make it nearly impossible for anyone who is not affiliated with an already well-established academic institution or business to contribute (Strubell et al., 2019; Schwartz et al., 2019). The appalling lack of diversity in the field (Snow, 2018) means the necessary resources to take part in the machine learning community are overwhelmingly available to those who are wealthy, white, male, or able-bodied at the expense of those who do not match most of these criteria. Thus, the elements that drive AI's large carbon footprint also play into social effects.

Research on the environmental impacts of AI offers both short-term and long-term suggestions to make AI less carbon intensive. Methodologies and frameworks offer immediate ways such that researchers can assess, understand, and mitigate their environmental impacts, while slower, cultural changes to how AI is developed are meant to take shape gradually. Prior research acknowledges the need to balance environmental goals with the importance of ground-breaking research and innovation. We believe this is indeed important, considering the non-negligible (environmental and otherwise) benefits we can get from AI. However, two issues arise:

First, it is unclear who gets to decide what potentially innovative research is seemingly valuable enough to be pursued, and according to what criteria. Considering that limiting energy use and carbon footprint can place non-negligible constraints on a project (even though this may be reasonable in light of a cost-benefit analysis), in-depth discussions will undoubtedly be necessary to determine when research should or should not be done at the detriment of the environment. We believe that embodying a participatory design (Gupta and De Gasperis, 2020) approach in making these decisions will ultimately help develop technology that is not only aligned with the norms and values of the community that the technology seeks to serve but also to create greater accountability and transparency in the process.

Second, if a significant portion of AI research continues to strive primarily for accuracy over energy efficiency and curbing environmental harm, will AI research that focuses on efficiency be seen as "second class" AI? Energy-efficient AI may be less prestigious because it may not attain the same levels of accuracy and performance as AI that is unrestricted in how much energy it uses. As a result, AI focused on efficiency may be seen by many at the top of the field as inferior to AI research centered on performance and accuracy. Nothing short of an overhaul of the AI community culture seems necessary to avoid this. This idea of a focus on single metrics (Thomas & Uminsky, 2020) to make design tradeoffs has been shown to be detrimental to the development of technology and adopting a more comprehensive approach that can internalize some of the externalities presents a great starting point to address this problem.

*How this project is different from existing work*

The few studies done at the intersection of AI and environmental issues have made important contributions to achieving a comprehensive and standardized method for calculating carbon impacts. Of course, arriving at such a result is an iterative and collaborative process. And disagreements in terms of what elements to include are often productive and foster innovation. It is in this spirit that we undertake this research work.

To begin with, let us consider how each group of researchers attempts to measure energy use and carbon impact. Strubell et al. (2019), Lacoste et al. (2019), and Henderson et al. (2020) agree on calculating carbon intensity (CO2eq per kWh) as a way to measure the environmental impact of AI models. Schwartz et al. (2019) argue that measuring floating point operations (FPOs) is a more accurate way of assessing energy use and subsequent environmental impact. However, Henderson et al. (2020) claim that FPOs are not as reliable as some claim to measure energy use. This, among other issues, leaves us in a confusing situation as to how energy use and environmental impact should be measured for AI. We plan to investigate these discrepancies in order to propose a sound methodology that considers each paper's position.

It is important to note that Henderson et al. have taken notice of the lack of standardization in the work being done on the environmental impacts of AI, and attempt to remediate it by offering a standardized approach that takes into account the work by Strubell et al., Lacoste et al., and Schwartz et al. However, we believe some important elements are left out of Henderson et al.'s framework. Hence, we have a similar aim of offering a standardized framework for understanding and mitigating the environmental harm caused by AI, but hope to offer a more comprehensive and applied approach which is ultimately necessary for widespread adoption and use of the measure.

Interestingly, many of the elements we had outlined to accomplish prior to diving in the existing research (the standardized approach, the social badge for green AI, the technical innovations) are mentioned by Henderson et al. as important goals and promising avenues to make AI less carbon intensive. We dare hope that these similarities in our thinking highlight how useful an approach centred around these elements could be.

One area we aim to explore further is the effectiveness of carbon offsets as well as big companies' claims surrounding the use of renewable energy. Lacoste et al. (2019) and Henderson et al. (2020) engage critically with cloud providers' claims about carbon neutrality. Schwartz et al. (2019) briefly show skepticism towards claims made about the efficiency of carbon offsets, while Strubell et al. (2019) highlight two caveats to carbon offsets and renewable energy use. Critical engagement with these claims is essential for a few reasons. First, big tech companies have much to gain from customers perceiving them as environmentally friendly and as at the edge of innovation in terms of renewable energy use. This leads to perverse incentives and potential for "green-washing". Presenting themselves as such is attractive to the growing number of individuals who care about (or at least, want to be perceived as caring about) environmental issues and climate change. Second, in comparing carbon offsets to other means of curbing carbon emissions (like reducing overall energy use), it may be the case that carbon offsets are the most effective. This could significantly affect the best way forward for the development of environmentally sensible AI systems.

Strubell et al. (2019), Lacoste et al. (2019), and Henderson et al. (2020) address the location of the power grid on which one's AI model is trained in terms of carbon emissions, there seems to be agreement concerning how central this feature can be in curbing one's AI research carbon emissions. Henderson et al. (2020) include the most parameters because, according to them, overly simple estimates of carbon emissions are imprecise and can significantly under- or overestimate the amount of carbon emitted.

**ESG framework**

Our SECure Environmental, Social, and Governance (ESG) framework targets different audiences in varying ways. To begin, the first pillar of compute-efficient machine learning, primarily targets AI researchers and practitioners. To a certain extent, compute-efficient machine learning can potentially have a large social impact in terms of access to the means necessary to do AI research. Indeed, greater efficiency in terms of compute needed could drastically lower the barrier to entry for individuals like undergraduate researchers (Schwartz et al., 2019) and/or those who are not affiliated with wealthy universities and organizations (Strubell et al. 2019). This is because, for one, the hardware needed to train an AI model is currently very expensive, and while cloud-based servers are cheaper, they still do not allow "any inspired undergraduate with a laptop has the opportunity to write high-quality papers that could be accepted at premier research conferences" (Schwartz et al., 2019). If AI is more compute-efficient to the point where it requires only a laptop or other relatively obtainable hardware, the field of AI may become much more accessible. Compute-efficient machine learning could thereby have a sizable social impact.

The second and third pillars, federated learning and data sovereignty, directly target AI researchers and practitioners. Both are primarily aimed at AI practitioners and researchers because, once again, these are techniques to be implemented by individuals in these positions. However, in a secondary manner, these are also addressed to customers, as they tend to value privacy, and welcome more secure data analysis for AI. Presented with two options, one less secure and one more secure, it is reasonable to expect, all else being equal, that consumers will choose the most secure option. In this case, this is represented by the pillars of federated learning and data sovereignty.

The fourth and last pillar, the LEEDesque certificate, targets consumers as well as the AI industry. The certificate is an opportunity for consumers to choose environmentally sensible AI. This means the industry may now have some added economic incentive to limit unnecessary environmental impact. Change in customer behaviour (and subsequently, change in industry behaviour) may happen through, for instance, social pressure (Mani et al., 2013) related to making a choice (i.e. purchasing environmentally sensible AI systems) that is associated with a more virtuous and positive outcome (e.g. helping curb carbon emissions, which can help slow climate change). A public display of using solutions that carry such a certification of mark is a signal in one's social circles of being well-intentioned and taking one's civic duties seriously. Prior success with these virtue signals has shown to make a shift in industry norms as seen in food products consumption that follows environmental best practices and the electric-vehicle industry.

Components of the SECure framework

1. Compute-efficient ML

Using compute-efficient machine learning methods has the potential to lower the computation burdens that typically make access inequitable for researchers and practitioners who are not associated with large organizations that have access to heavy compute and data processing infrastructures. As an example, recent advances in the quantization of computations in neural networks (Jacob et al., 2018) have led to significant decreases in computational requirements. This also allows for the use of lower-cost resources like CPUs compared to more expensive hardware for the training of complex architectures which typically require GPUs or TPUs.

Studies such as the one by Jouppi et al. (2017) highlight the performance tradeoffs and give an indication on a pathway to incorporating hardware improvements such as the use of specialized chips, ASICs (Application Specific Integrated Circuits), for machine learning related computations. Though, we see the access and limited availability of such hardware as a potential barrier, the possibility of cost-efficiency makes this approach promising. Documenting the use of specific underlying hardware for the training of systems within the framework paired with benchmarking of performance metrics will provide one piece of essential information in the computation of the final metric.

Another area of ML research that has bearing for compute-efficient machine learning is that of machine learning models for resource-constrained devices like edge-computing on IoT devices. For example, with devices that have RAM sizes in KB, model size can be minimized along with prediction costs using approaches like Bonsai (Kumar et al., 2017) that proposes a shallow, sparse tree-based algorithm. Another approach is called ProtoNN that is inspired by kNN but uses minimal computation and memory to make real-time predictions on resource-constrained devices (Gupta et al., 2017). Novel domain-specific languages like SeeDot (Gopinath et al., 2019), which expresses ML-inference algorithms and then compiles that into fixed points, makes these systems amenable to run on edge-computing devices. Other distilled versions of large-scale networks like MobileNets (Howard et al., 2017) and the growing prevalence of TinyML will also bring about cost- and compute-efficiency gains.

This part of the framework proposes the computation of a standardized metric that is parametrized by the above components as a way of making quantified comparisons across different hardware and software configurations allowing people to make informed decisions in picking one solution over another. We are currently in the experimental phases to assess the right formulation capturing these statistics into a mathematical equation that enables a comprehensive comparison from the hardware and software configuration standpoint.

2. Federated learning

As a part of this framework, we propose the utilization of federated learning (Bonawitz et al., 2019) approaches as a mechanism to do on-device training and inference of ML models. The purpose of utilizing this technique is to mitigate risks and harm that arises from centralization of data, including data breaches and privacy intrusions. These are known to fundamentally harm the trust levels that people have in technology and are typically socially-extractive given that they may use data for more than the purposes specified when the data is sourced into a single, centralized source. Federated learning also has the second-order benefit of enabling computations to run locally thus potentially decreasing carbon impacts if the computations are done in a place where electricity is generated using clean sources. We acknowledge that there may be gains to be had from an "economies of scale" perspective when it comes to energy consumption in a central place—like for a data center that relies on government-provided access to clean energy. This is something that still needs to be validated, but the benefits in terms of reducing social harm are definite, and such mechanisms provide for secure and private methods for working on data that constitutes personally identifiable information (PII).

Our goal with this research work is to empirically assess these methods and provide information to the developers such that they can adopt these mechanisms. We also aim to empower users to demand such solutions rather than resign to technology-fatalism.

3. Data sovereignty

Data sovereignty refers to the idea of strong data ownership and giving individuals control over how their data is used, for what purposes, and for how long. It also allows users to withdraw consent for use if they see fit. In the domain of machine learning, especially when large datasets are pooled from numerous users, the withdrawal of consent presents a major challenge. Specifically, there are no clear mechanisms today that allow for the removal of data traces or of the impacts of data related to a user in a meaningful manner from a machine learning system without requiring a retraining of the system. Preliminary work (Bourtoule et al., 2019) in this domain showcases some techniques for doing so—yet, there is a lot more work needed in this domain before this can be applied across the board for the various models that are used.

Thus, having data sovereignty at the heart of system design which necessitates the use of techniques like federated learning is a great way to combat socially-extractive practices in machine learning today.

Data sovereignty also has the second-order effect of respecting differing norms around data ownership which are typically ignored in discussions around diversity and inclusion as it relates to the development of AI systems. For example, indigenous perspectives on data (Kukutai & Taylor, 2016) are quite different and ask for data to be maintained on indigenous land, used and processed in ways that are consistent with their values. This is something that can be captured more holistically which is why we include it as a part of the SECure framework. The precise incorporation of this into the framework will depend on the research that is carried out as a part of this work.

4. LEEDesque certification

The certification model today relies on some sort of a trusted, third-party, independent authority that has the requisite technical expertise to certify the system meets the needs as set out in standards, if there are any that are widely accepted. Certificates typically consist of having a reviewer who assesses the system to see if it meets the needs as set out by the certifying agency. The organization is then issued a certificate if they meet all the requirements. An important, but seldom discussed component of certification is something called the Statement of Applicability (SoA).

Certificates are limited in terms of what they assess. What the certifying agency chooses to evaluate, and the inherent limitation that these choices are representative of the system at a particular moment in time with a particular configuration. This is typically addressed—what gets left out of the conversation is the SoA and how much of the system was covered under the scope of evaluation. The SoA is also not publicly or easily available, while the certification mark is shared widely to signal to consumers that the system meets the requirements as set out by the certification authority. Without the SoA, one cannot really be sure of what parts of the system were covered. This might be quite limiting in a system that uses AI, as there are many points of integration as well as pervasive use of data and inferences made from the data in various downstream tasks.

*What are some best practices to make certificates more effective?*

Recognizing some of the pitfalls in the current mechanisms for certification, our proposal is for the certification body to bake in the SoA into the certificate itself such that there is not a part of the certification that is opaque to the public. Secondly, given the fast-evolving nature of the system, especially in an online-learning environment for machine learning applications, we see the certificate having a very short lifespan. An organization would have to be recertified so that the certificate reflects as accurately as possible the state of the system in its current form.

Certification tends to be an expensive operation and can thus create barriers to competitiveness in the market where only large organizations are able to afford the expenses of having their systems certified. To that end, we require that the certification process be automated as much as possible to reduce administrative costs—as an example, having mechanisms like Deon (*About — Deon*, n.d.) might help. Also, tools that would enable an organization to become compliant for a certification should be developed and made available in an open-source manner.

*Standardized measurement technique*

The proposed standardization will also serve to allow for multiple certification authorities to offer their services, thus further lowering the cost barriers and improving market competitiveness while still maintaining an ability to compare across certificates provided by different organizations. An additional measure that we have deemed to be of utmost importance is to have the certificate itself be intelligible to a wide group of people. It should not be arcane and prevent people from understanding the true meaning and impact of the certification. It will also empower users to make choices that are well-informed.

*Survey Component*

To build on the point made above, the goal of the certification process is to empower users to be able to make well-informed choices As a part of this research work, we will be embarking on extensive user survey to identify what information users are seeking from certification marks and how that information can be communicated in the most effective manner.

Additionally, triggering behaviour change on the part of the users through better-informed decisions on which products/services to use needs to be supplemented with behaviour change on the part of the organizations building these systems. We believe that clear comparison metrics that allow organizations to assess the state of their systems in comparison with actors in the ecosystem will also be important. Keeping that in mind, a survey of the needs of practitioners will help ensure the certification is built in a manner that meets their needs head-on, thereby encouraging widespread adoption.

*Data storage and Usability*

Software developers and Machine Learning (ML) engineers work with data files that are not easy to use among general audiences that lack programming experience. For example, JSON is a common file format that is used by developers when working with and analyzing web data. JSON is efficient for storing massive amounts of nested data. In this format, data takes up less machine storage space than more user-friendly file formats, such as Excel or CSV. While JSON is more efficient in terms of compute storage and perhaps memory, therefore environmentally efficient; ML engineers do not often work in isolation. In corporate settings, developers work in or collaborate with data science departments. This means that it is often necessary to convert files to formats that are usable to those without computer programming skills, such as Excel or CSV file formats. This conversion is very costly. For example, approximately 250 MB of JSON data or less, as a CSV file, converts to a file size that is over 500 MB. Converting JSON to CSV, in this instance, at least doubles the file size and need for machine storage. This example does not account for the memory or compute power required to make the conversion. In isolation, or in individual workflows, file conversion tasks may seem less computationally demanding than running an image classification model on the cloud, for example; but, added up, these tasks across developers and organizations significantly increase the environmental cost of computation. Importantly, file conversion tasks are avoidable, if we design user-interfaces for data and data science work that are usable by a non-programming audiences, and make it so that user-interfaces for data and data science do not require efficiently stored data to be converted to inefficient formats.

In our work, we propose to measure the cost of file conversion work, both in terms of storage and compute power (memory). We will integrate our cost estimate models into the software packages we are developing. These software tools would allow developers to estimate the environmental cost of each development or engineering task, in real time. These real-time estimates would allow developers to observe the efficiency or cost of each data task and whether their workflows could be designed in a more environmentally friendly way.

**Future research directions**

The potential future research that may follow from this project could contribute significantly to making AI research more accessible as well as more environmentally sensible. From a broad perspective, this project lends itself well to future recommendations in terms of public policy. One could devise a framework to create public compute facilities that make it easier for people who are not affiliated with large organizations to work on AI applications. In addition, inquiring into making this as cost- and energy-efficient as possible while ensuring it remains accessible and powerful enough to foster quality research appears crucial to us. To accompany public compute facilities, a data commons (Miller et al., 2008) could also be useful, and has the possibility of making large amounts of quality data more accessible to researchers while upholding individuals' privacy. Particularly in a supervised machine learning setting, it is important to have high-quality data to do a meaningful analysis. Data co-operatives (Hafen et al., 2014) are another solution in this domain that if implemented in a practical fashion and adopted widely will lead to more equitable outcomes and bring about inclusion for people who are currently marginalized. Another avenue for exploration is to investigate the use of small data approaches and meta-learning that have the likelihood of being more inclusive by minimizing the need for extensive data collection for making predictions and doing classification.

Given the strong influence that market forces have on which solutions are developed and deployed, we see the SECure certificate as a mechanism creating the impetus for consumers and investors to demand more transparency on the social and environmental impacts of these technologies and then use their purchasing power to steer the progress of development in this field that accounts for these impacts. Responsible AI investment, akin to impact investing, will be easier with a mechanism that allows for standardized comparisons across various solutions, which is what SECure is perfectly geared towards.

**Conclusion and Final Remarks**

In this paper, we have presented a novel framework titled SECure that combines elements of social and environmental impacts into a single instrument to enhance decision-making when it comes to the development and deployment of AI-enabled solutions in a responsible fashion. We laid out the groundwork for the importance of considering both the environmental and social impacts, and how this has the potential to democratize access to AI development and use. We also explored how these considerations can lead to solutions that are more inclusive. Expanding on the details of our framework, we review the most pertinent approaches that will be required to make SECure a comprehensive evaluation including the approaches of compute-efficient machine learning, federated learning, data sovereignty, and a LEEDesque certification. We also expand on the essential features of this certification to enable developers, consumers, and investors to make informed decisions. Finally, we conclude with how this research work will lay the groundwork for future efforts in helping us build responsible AI systems in a more concrete manner.